Correspondence and requests for materials should be addressed to Q. H. G. (qhgong@pku.edu.cn) and X. Y. H. (xiaoyonghu@pku.edu.cn)

# Integrated all-optical logic discriminators based on plasmonic bandgap engineering

Cuicui Lu, Xiaoyong Hu, Hong Yang & Qihuang Gong

State Key Laboratory for Mesoscopic Physics & Department of Physics, Peking University, Beijing 100871, People's Republic of China

**Abstract**

Optical computing uses photons as information carriers, opening up the possibility for ultrahigh-speed and ultrawide-band information processing. Integrated all-optical logic devices are indispensible core components of optical computing systems. However, up to now, little experimental progress has been made in nanoscale all-optical logic discriminators, which have the function of discriminating and encoding incident light signals according to wavelength. Here, we report a strategy to realize a nanoscale all-optical logic discriminator based on plasmonic bandgap engineering in a planar plasmonic microstructure. Light signals falling within different operating wavelength ranges are differentiated and endowed with different logic state encodings. Compared with values previously reported, the operating bandwidth is enlarged by one order of magnitude. Also the SPP light source is integrated with the logic device while retaining its ultracompact size. This opens up a way to construct on-chip all-optical information processors and artificial intelligence systems.

## Introduction

Ultrahigh-speed and ultrawide-band information processing based on optical computing is a long-term pursuit of researchers. Adopting photons as information carriers can overcome the inherent performance limitations of modern semiconductor-based electronic central processing units (CPUs) such as serious thermal losses and poor processing rates[1,2]. Integrated all-optical logic devices are key and core components of photonic CPUs in optical computing systems. Unfortunately, to date, there are few experimental achievements in nanoscale all-optical logic devices suitable for on-chip integration applications[3-6]. Nanoscale all-optical logic discriminator functions to discriminate and encode incident light signals according to wavelength, which forms the essential basis of logic processing systems of photonic CPU chips[7]. The traditional realization of all-optical logic discriminator was based on the optical fiber Bragg grating, the large size of which makes it very difficult for practical on-chip integration applications[8-10]. Moreover, the operating bandwidth attained was as small as tens of nanometers[8-10]. In deed, to date, a nanoscale integrated all-optical logic discriminator with an ultrawide operating bandwidth has not been reported. This has hindered practical applications of all-optical logic discriminator.

Here, we report a nanoscale all-optical logic discriminator suitable for on-chip integration applications, realized by implementing plasmonic bandgap engineering in a planar plasmonic microstructure. As a kind of electromagnetic mode coupled to the oscillations of conduction electrons that propagate along the interface between metal and dielectric materials, surface plasmon polaritons (SPPs), has been regarded as a promising candidate to significantly miniaturize all-optical devices because of its unique subwavelength field confinement properties[11]. The incident light signal is converted into SPPs by a nanoslit, which acts as a photon-to-SPP convertor widely used in the fields of plasmonics and nanophotonics[12,13].

Adopting a continuous wave (CW) laser system with a line width of 1.5 nm as the light source, only the quasi-monochromatic SPP specified can be generated by the nanoslit under excitation of the CW light signal. This has been confirmed by our previous experiments[5,6]. A one-dimensional (1D) plasmonic crystal, constructed by spatially periodic arrays of air grooves etched in the metal surface, has remarkable SPP stop band owing to the strong Bragg scattering effect[14,15]. Moreover, the frequency position of the edge of the stop band can be changed by adjusting the structure parameters of the plasmonic crystal, i.e., through plasmonic bandgap engineering[16,17]. When the resonant wavelength of the quasi-monochromatic SPP fall within the stop band of the plasmonic crystal, the SPP will be reflected completely by the plasmonic crystal, and subsequently, no output signal can be obtained in the output port. This corresponds to logic state "0". In contrast, when the resonant wavelength of the SPP mode falls within the pass band, the SPP mode propagates through the plasmonic crystal. A strong signal can then be obtained in the output port, corresponding to logic state "1". With plasmonic bandgap engineering, it is possible to produce 1D plasmonic crystals with different structure parameters having different cutoff wavelengths (corresponding to the edge of the stop band). Therefore, when three different plasmonic crystals are arranged in a row, the visible and near-infrared range will be divided into different wavelength regions. The incident light signals falling within different wavelength regions can be endowed with different logic state encodings. Compared with previously reported bandwidths, the operating bandwidth is enlarged by one-order of magnitude, and the SPP light source is integrated with the logic device. Moreover, the planar configuration and the ultracompact device size make the all-optical logic discriminator device much more suited for practical integrated logic chip applications.

**Results**

The schematic cross-section structure of a single logic discriminator unit is shown in Fig. 1a. Each unit is composed of a nanoslit connected with a deeply-etched 1D plasmonic crystal on the left-hand side and a shallow-etched 1D plasmonic crystal on the right-hand side in a 300-nm-thick gold film coated with a 150-nm-thick organic polymer layer made of polyvinyl alcohol (PVA) on silicon dioxide substrate. All the air grooves in the sinistral (or the dextral) 1D plasmonic crystals are identical and have the same structural parameters. The air grooves do not need individually shaping. The nanoslit and the sinistral 1D plasmonic crystal constitute a unidirectional SPP launcher, which converts incident signal light into SPPs, and subsequently launches the SPPs towards the right along the interface between gold film and PVA layer. To completely prohibit the SPPs from propagating towards the left, the sinistral plasmonic crystal is etched deeply. The dextral plasmonic crystal performs the function of distinguishing different light signals based on the stop band effect. The top-view scanning electron microscopy (SEM) image of the logic discriminator sample is shown in Fig. 2b. There are three discriminator units, labeled units 1, 2, and 3. The nanoslits in all units have the same structural parameters: depth 450 nm, width 200 nm, and length 8 μm. The sinistral plasmonic crystals in units 1, 2, and 3 also have the same structural parameters. Each sinistral plasmonic crystal has seven air grooves of depth 220 nm, width 200 nm, and length 8 μm; the lattice period of each is 340 nm. The distance between the nanoslit and the first air groove of the sinistral plasmonic crystal is 340 nm for all units. Although there are six air grooves with depth 150 nm and length 8 μm for the dextral plasmonic crystals in all units, other structural parameters associated with these dextral plasmonic crystals are different. The air groove width of the dextral plasmonic crystal is 140 nm for unit 1, 160 nm for unit 2, and 180 nm for unit 3. The lattice period for each dextral plasmonic crystal is 280 nm for unit 1, 320 nm for unit 2, and 360 nm for unit 3. The distance between the nanoslit and the first air groove of these dextral plasmonic crystals is 280 nm for unit 1, 320 nm for unit 2, and 360 nm for unit 3.

Two identical decoupling gratings were also deeply etched in the output ports of each unit to convert SPPs into scattered light for the purpose of measurement. Every decoupling grating had three air grooves of width 260 nm, depth 250 nm, and length 6 μm. The lattice period of the decoupling gratings was 520 nm. The distance between the nanoslit and decoupling gratings on both sides was 8 μm. Under excitation of an incident CW light signal, with a limited and small line width of 1.5 nm, the generated quasi-monochromatic SPP at the bottom of nanoslit propagated upward along two sidewalls, and then along the interface between the gold film and PVA layer. To clarify the underlying physical mechanisms of the all-optical logic discriminator, we describe how the etched plasmonic nanostructure can modulate the propagation properties of the different SPP modes propagating along the interface. Krasavin *et al.* have pointed out that if the gold film is thicker than 100 nm in the thick dielectric substrate/gold/isolator microstructure, the coupling between the SPPs in the substrate/gold interface and in the gold/isolator interface is too weak, hence the effects from the thick dielectric substrate on SPPs propagating along the gold/isolator interface can be neglected[18]. The modal structure of the glass/Au/PVA/air geometry can also support a waveguide-like mode in the PVA layer. However, this mode can not be excited directly by the incident light signal through the nanoslit because the wavevector of the external incident light signal mismatches that of the waveguide-like mode, which has been confirmed by Wu's calculations[19].

**Influences of PVA layer on SPP propagation properties.**

To study the influences of the 150-nm-thick PVA layer on the propagation properties of SPP modes, we calculated the SPP propagation length as a function of incident light wavelength in a 300-nm-thick gold film coated with a 150-nm-thick PVA layer using the finite element method (adopting a commercial software package COMSOL Multiphysics); results are shown in Fig. 1c. The wavelength-dependent complex dielectric function of gold was obtained from Ref. 20. The refractive index of PVA was 1.5 in the visible and near infrared range[21]. Even SPPs at short wavelength can still propagate a couple of microns untill their intensity is smaller by factor 1/e, as shown in Fig. 1(c), which ensures that SPPs reach the decoupling gratings, and strong scattered signals can be obtained at the decoupling gratings. To further clarify the role of the 150-nm-thick PVA layer, we calculated using the finite element method the magnetic-field distribution in the plane of 75 nm above the gold film for two structures: an 800 nm incident light in Au/air (300-nm-thick gold film in air) structure and an Au/PVA/air (300-nm-thick gold film coated with a 150-nm-thick PVA layer in air) structure; results are shown in Fig. 1d. There was a small magnetic-field distribution for the SPP mode in this plane for the Au/air structure, but the distribution was different for the Au/PVA/air structure, i.e., the structure used in our experiment. There was strong magnetic-field distribution associated with the SPP mode in the PVA layer, which implies that a very tight field confinement is formed in the Au/PVA/air structure. This is further confirmed by our calculated power flow distribution (Fig. 1e) in the Y direction for the SPP mode, with an origin at the bottom of gold film and 600 nm away from the nanoslit, for the Au/air and Au/PVA/air structures. About 82% of the power flow was confined in the PVA layer for the Au/PVA/air structure. Thus a much stronger modulation of the SPP propagation is produced despite only several nanogrooves being etched in the Au/PVA/air structure compared with that in the Au/air structure; this has been confirmed by Chen's measured results[22]. Thus, an ultrasmall feature size of the all-optical logic device can be anticipated.

**Wide-band unidirectional SPP launching properties.**

The nanoslit and the sinistral 1D plasmonic crystal constitute a unidirectional SPP launcher. The sinistral 1D plasmonic crystal can produce a remarkable SPP stop band if the lattice period is nearly half the SPP wavelength[14]. Thus, the sinistral plasmonic crystal acts as a Bragg reflector. The air groove depth of the sinistral plasmonic crystal is 220 nm, indicating that the air grooves are etched through to the PVA layer and also etched into the gold film. Because of the deeply etched air grooves, the Au/PVA/air structure modulated SPP propagation enormously. The left-going SPPs for which their wavelengths fall within the stop band of the sinistral plasmonic crystal are reflected completely, and then propagate towards the decoupling grating at the far right-hand side. To study the unidirectional SPP launching properties, we fabricated a SPP launcher (Fig. 2a) consisting of a nanoslit connected with a 1D plasmonic crystal on the left-hand side, having the same structural parameters as that of the logic discriminator sample. A single isolated nanoslit was also etched in the top side to serve as an on-chip reference. For the purpose of measurement, two identical decoupling gratings on the left- and right-hand sides of the SPP launcher were used to decouple the SPP modes to free space. The calculated transmission spectrum of the sinistral plasmonic crystal as a function of incident light wavelength using the finite element method (Fig. 2b) clearly shows that there is an ultrawide-band SPP stop band from 600 nm to 1050 nm, covering a wavelength range of 450 nm. To experimentally verify the calculated results, we measured the CCD images of the SPP launcher sample under excitation of a p-polarized CW laser with the magnetic field vector parallel to the nanoslit in the Z direction. Limited by the tuning wavelength range of the Ti:sapphire laser system, we only measured the SPP launcher sample under excitation of different wavelengths ranging from 750 to 1000 nm; measured results are shown in Fig. 2c-h. In excitations of the isolated nanoslit, strong scattered light was observed from two decoupling gratings on both the left- and right-hand sides. SPPs were generated almost equally on both sides of the nanoslit. Thus, left- and right-going SPPs simultaneously existed because of the symmetric configuration of the isolated nanoslit. That the intensities at the decoupling gratings on both sides were not completely identical is caused by the imperfect etching. A different result arises if the nanoslit connected with a plasmonic crystal is excited. Strong scattered light is observed only from the decoupling grating in the right-hand side if the incident light wavelength lies between 750 and 1000 nm. No scattered light is seen from the decoupling grating on the left-hand side. This implies that there exists only right-going SPPs, and the SPPs launched leftward are reflected completely. Unidirectional SPP launching is attained in a very wide wavelength range from 750 to 1000 nm. The operating bandwidth, 250 nm, is enlarged by two orders of magnitude compared with those previously reported[13,21,22]. We also measured the launching efficiency ratio as a function of incident laser wavelength; results are shown in Fig. 2i. The launching efficiency ratio is defined as the quotient between the intensities of the right-going SPPs with and without the 1D plasmonic crystal. In our experiment, the launching efficiency ratio is obtained as the quotient between the scattered intensities of right-hand side decoupling gratings with and without the sinistral 1D plasmonic crystal. The scattered intensities are extracted from the measured CCD images. The average SPP launching efficiency ratio was measured at 1.78, which is higher than those previously reported[13,21,22]. Moreover, no serious oscillations appeared in the measured signal curve. This implies that there is no destructive interference between the reflected SPP modes and the original right-going SPP modes. The distance between the nanoslit and the first air groove of the sinistral plasmonic crystal was 340 nm, corresponding to the lattice period of the sinistral plasmonic crystal. Therefore, the left-going SPPs will be directly reflected, and subsequently, constructively interfere with the originally right-going SPPs. Thus, a wideband unidirectional SPP launching has been produced using a simple configuration.

**All-optical logic discriminator performance.**

To study the function of the all-optical logic discriminator sample, we first calculated using the finite element method the transmission spectra of the 1D plasmonic crystal in the right-hand side of the nanoslit as a function of the incident laser wavelength; results are shown in Fig. 3a. Perfect SPP stop bands appear for the dextral plasmonic crystals of all units despite the air grooves being only etched through to the 150-nm-thick PVA layer. A strong magnetic-field distribution for SPP modes (Fig. 1d) is tightly confined in the PVA layer for the Au/PVA/air structure. Therefore, when a 1D plasmonic crystal is etched through the cover PVA layer, a stop band for the guided SPP mode will be formed if the lattice period is nearly half the SPP wavelength; this has been confirmed by our previous measurements[23]. The position of the edge of the stop band in the long-wavelength direction increases gradually for dextral plasmonic crystals of all units. Gan *et al.* have pointed out that the dispersion relations of a 1D plasmonic crystal can be tuned by adjusting the air groove width or lattice period, shifting the stop band edge accordingly[16]. For the 1D plasmonic crystal constructed from a gold film-coated dielectric grating, the stop band edge shifts to shorter wavelengths with increasing air groove width, whereas the stop band edge shifts to longer wavelengths with increasing lattice period; this has been confirmed by Gan's measured results[17]. In our experiment, there is a 20-nm increment in the air groove width of the dextral plasmonic crystals, viz, 140, 160, and 180 nm for units 1, 2, and 3, respectively. This makes the stop band edge shift to shorter wavelengths. However, an even larger increment, 40 nm, in the lattice period of dextral plasmonic crystals occurred, viz, the lattice periods of 280, 320, and 360 nm for dextral plasmonic crystals of units 1, 2, and 3, respectively. This makes the stop band edge shift to longer wavelengths. According to our calculation, the magnitude of the red-shift was much larger than that of the blue-shift. As a result, the stop bandedge shifts to longer wavelengths. By plasmonic bandgap engineering, i.e., carefully choosing the structural parameters of three dextral 1D plasmonic crystals, the positions at the long-wavelength edge of the stop band were 770, 830, and 887 nm for the dextral plasmonic crystals of units 1, 2, and 3, respectively. As a result, the visible and near-infrared range was divided into four wavelength regions, namely, the Green region (ranging from 650 to 770 nm), the Cyan region (ranging from 770 to 830 nm), the Yellow region (ranging from 830 to 887 nm), and the Orange region (ranging from 887 to 1000 nm), as shown in Fig. 3a. The Green region is located in the stop band of the dextral plasmonic crystals of all units; the Cyan region is situated in the pass band of the dextral plasmonic crystal of unit 1, and in the stop bands of dextral plasmonic crystals of units 2, and 3; the Yellow region is situated in the pass bands of dextral plasmonic crystals of units 1 and 2, and in the stop band of the dextral plasmonic crystal of unit 3; the Orange region is within the pass bands of dextral plasmonic crystals of all units. As a result, if two incident light signals fall within the same wavelength region, all units impose the same modulation on SPP propagation. In contrast for two incident light signals falling within different wavelength regions, the modulation changes for at least one unit, from blocking to conducting, or vice versa.

To experimentally confirm the all-optical logic discriminator functions, we took CCD images of the logic discriminator sample under excitation of different CW lasers; results are shown in Fig. 3b-g. With a wavelength of 750 nm (falling within the Green region), no scattered light can be obtained from decoupling gratings in the output ports (Fig. 3b), corresponding to output logic state encoding of "000". Under excitation at the 750-nm wavelength, the wavelength of the generated quasi-monochromatic SPP mode is located in the stop bands of dextral plasmonic crystals of all units. Thus, the SPP mode is reflected completely by three dextral plasmonic crystals, and is unable to reach the decoupling gratings. With wavelength of 800 nm (falling within the Cyan region), strong scattered light was obtained only from the decoupling grating in the output port of unit 1 (Fig. 3c); no scattered light can be obtained from decoupling gratings in the output ports of units 2 and 3. This corresponds to an output logic state encoding of "100". Under excitation at the 800-nm wavelength, the wavelength of the generated quasi-monochromatic SPP mode is located in the pass band of the dextral plasmonic crystal of unit 1, and within the stop bands of the dextral plasmonic crystals of units 2 and 3. As a result, the SPP mode can only propagate through unit 1, and subsequently arrived at the decoupling grating of unit 1. With wavelength of 850 nm (falling within the Cyan region), strong scattered light was obtained only from the decoupling gratings in the output ports of units 1 and 2 (Fig. 3d). No scattered light can be obtained from the decoupling grating of unit 3. The corresponding encoding is "110". Under excitation at 850-nm wavelength, the wavelength of SPP mode is located in the pass band of the dextral plasmonic crystals of units 1 and 2, and within the stop band of the dextral plasmonic crystal of unit 3. As a result, the SPP mode can propagate through the dextral plasmonic crystals of units 1 and 2, and subsequently arrived at the decoupling grating positions of units 1 and 2. With wavelength of 900 nm (falling within the Orange region), strong scattered light is obtained from the decoupling gratings in the output ports of all units, as shown in Fig. 3e. This corresponds to an output logic state encoding of "111". Under excitation at 900-nm wavelength, the wavelength of the SPP mode was located in the pass bands of dextral plasmonic crystals of all units. Again, the SPP mode can propagate through the dextral plasmonic crystals of all units, and subsequently arrived at the decoupling gratings of all units. With wavelength of 950 nm (also falling within the Orange region), strong scattered light is obtained from the decoupling gratings in the output ports of all units (Fig. 3f). Again, the output logic state encoding of "111" was achieved. The same case occurs for wavelength of 1000 nm (Fig. 3g). These evidences confirm the excellent performance of the all-optical logic discriminator.

## Discussion

To further confirm the function of the all-optical logic discriminator sample, we also calculated using the finite element method the power flow distribution at cross-sections of three logic discriminator units under excitation of different signal light wavelengths; results are shown in Fig. 4. Clearly, for incident 750-nm CW light, no SPP mode can reach the dextral decoupling grating positions of any unit

(Fig. 4a), yielding encoding "000". For incident 800-nm CW light, a strong power flow of SPPs can reach the dextral decoupling grating of unit 1, but no power flow appeared in the dextral decoupling gratings of units 2 and 3 (Fig. 4b), yielding encoding "100". For incident 850-nm CW light, a strong power flow can reach the dextral decoupling gratings of units 1 and 2, but not the dextral decoupling grating of unit 3 (Fig. 4c), thus yielding encoding "110". For incident 900-nm CW light, strong power flow can reach the dextral decoupling gratings of all units (Fig. 4d), thus yielding encoding "111". A similar outcome occurs for incident 950- and 1000-nm light (Figs. 4e and f). These calculated results are in agreement with the measured results. This further confirms the excellent performance of the all-optical logic discriminator.

Because of the great difficulty of depositing a thin gold strip in the cross-section of the 150-nm-thick organic PVA cover layer to increase the electrical conductivity, we did not obtain the SEM image of the cross-section of the focused-ion-beam (FIB) milled discriminator sample. However, the perfect SPP stop band properties of the 1D plasmonic crystals constructed from a 150-nm-thick PVA gratings coated with a 300-nm-thick gold film have been confirmed by our measured results. Moreover, the excellent stop band properties of the plasmonic crystal constructed from an organic-polymer-grating coated gold film have been confirmed by our previous experiment[23]. By carefully choosing the structural parameters of the decoupling gratings, the wavelength dependence of decoupling gratings can be reduced greatly within a certain wavelength range. Accordingly, an approximately equal decoupling efficiency can be maintained in the wavelength range; this has been confirmed by Chen's measurements[20]. According to our calculation, the value of decoupling efficiency of the decoupling gratings is nearly maintained in the wavelength range from 750 to 1000 nm. The measured transmission looks like as a function of incident light wavelength, and reproduces the broad stop band which is surprisingly flat, as shown in Fig. 2c-h. The intensities measured at the reference structure (only nanoslit and decoupling gratings), as given in Fig. 2c-h, show a strong variation, but should be more or less equal. The reason can be explained as follows: First, the etching precision of our FIB system was about 10 nm. Limited by the poor etching precision of the FIB system, the reference structure was not etched perfectly. Second, both the spectrum response and the imaging efficiency of our CCD device were not perfect and varied with wavelength in the wavelength range from 750 to 1000 nm. We also measured CCD images of the launcher sample under excitation of a CW laser at wavelength 532 nm; the measured result is shown in Fig. 2j. The SPP excited by 532-nm CW laser is located in the pass band of the sinistral 1D plasmonic crystal (Fig. 2b). For excitations of the isolated nanoslit, strong scattered light is observed from two decoupling gratings in both the left- and right-hand sides. Left- and right-going SPPs simultaneously exist because of the symmetric configuration of the isolated nanoslit. That the intensities at the decoupling gratings on both sides are not completely identical is due to imperfect etching. If the nanoslit connected with a plasmonic crystal is excited, strong scattered light is observed in the decoupling gratings on both left- and right-hand sides, signifying that both right- and left-going SPPs, and the SPPs excited by the 532-nm CW laser can propagate through the sinistral 1D plasmonic crystal. This further confirms the perfect SPP band gap properties of the sinistral 1D plasmonic crystal. The map between wavelength selection and logic state was realized based on plasmonic bandgap engineering, without any high power requirements. The traditional realizations are based on third-order optical nonlinearity or linear interference effect in optical fiber[24,25]. The former requires a high incident light intensity of $GW/cm^2$ because of the small third-order optical nonlinearity of conventional materials[24]; the latter results in large device sizes[25]. Therefore, compared with previous devices, an ultracompact device size and ultralow operating power has been achieved simultaneously for the all-optical logic discriminator. The ultrawide-band unidirectional SPP light source and the all-optical logic discriminator have been integrated into a planar plasmonic microstructure, overcoming the prevailing difficulty that a nanoscale light source can not be integrated with all-optical logic devices. For previously reported all-optical logic devices, no on-chip light source has been integrated with any logic device[3-6]. The lateral dimensions of the patterned area of a single logic discriminator unit are far less than 4 μm, indicating that an ultracompact device size can be attained. The SPP modes excited by the incident light propagate along the interface between gold film and PVA layer. The planar configuration of the all-optical logic discriminator device is very suited to practical on-chip integration applications. Moreover, the operating wavelength range can be shifted very easily to the optical communication range by adjusting structural parameters of the logic discriminator units. Although the enlarged bandwidth does not improve data transmission capacity, there are significant advantages for logic discriminator devices with enlarged bandwidths, such as better wavelength tolerance and broader operating wavelength range. This has been confirmed in the experiments by Ou *et al.*[9]. Gan *et al.* have pointed out that plasmonic gratings have great potential applications in actively tunable integrated photonic devices[17]. Many active functionalities can be realized using plasmonic gratings, such as ultrafast tunable photonic routers, ultrafast all-optical switches, ultrafast photonic buffers, and actively tunable wavelength-division multiplexing (WDM) devices[17,23,26].

In conclusion, we reported a new strategy of implementing plasmonic bandgap engineering to realize a nanoscale integrated all-optical plasmonic logic discriminator. Light signals falling within different operating wavelength ranges are differentiated and endowed with different logic state encodings. This ultracompact device is robust, free from environment impact, and more suited for practical on-chip applications. This might offer a way for constructing integrated plasmonic circuits and realizing ultrawide-band and ultrahigh-speed information processing and artificial intelligence systems based on plasmonic chips.

## Methods

**Sample fabrication**. The gold film was fabricated using a laser molecular beam epitaxy (LMBE) growth system (Model LMBE 450, SKY Company, China). The beam (wavelength 248 nm, a pulse repetition rate 5 Hz) output from an excimer laser system (Model COMPexPro 205, Coherent Company, USA) was used as the excitation light source. The beam is focused onto a gold target mounted on a rotating holder, 15 mm away from the silicon dioxide substrate. A typical energy density of the excitation laser is about 500 mJ/cm$^2$. The growth rate measured by a film thickness/rate monitor is about 0.01 nm/pulse. PVA powder with an average molecular weight of 30,000 (J&K company, China) is dissolved in de-ionized water with a weight ratio of 1:32. The spin coating method is used to fabricate the PVA layer on the surface of the gold films. A FIB etching system (Model Helios NanoLab 600, FEI Company, USA) is employed to prepare the patterns of the nanoslit and the 1D plasmonic crystal. The spot current of the ion beam was only 7.7 pA to improve the etching quality.

**Micro-spectroscopy measurement setup.** In our experiment, a micro-spectroscopy measurement system is used to measure the unidirectional SPP launching properties and the all-optical logic discriminator sample. The nanoslit is normally illuminated from the back side using a p-polarized CW Ti:sapphire laser (Model Mira 900F, Coherent Company, USA) with different wavelengths. The optical-thick gold film can prohibit the direct transmission of the incident laser beam. The incident laser beam is focused to a spot of radius of about 50 μm, ensuring uniform illumination of the entire nanoslit. The line width of the laser spectrum curve is about 1.5 nm which ensures that only the specified quasi-monochromatic SPP modes can be excited by the nanoslit. The SPP mode is scattered using two decoupling gratings in the output ports. The scattered light is collected by a long working distance objective (Mitutoyo 20, NA=0.58) and then imaged onto a charge coupled device (CCD).

## Acknowledgements

This work was supported by the National Key Basic Research Program of China under grants 2013CB328704 and 2014CB921003, the National Natural Science Foundation of China under grants 11225417, 61077027, 11134001, 11121091, and 90921008, and the program for New Century Excellent Talents in University.

## Author contributions

C.L. and Q.G. proposed the idea. C.L., X.H., and H.Y. performed measurements. C.L., X.H., H.Y., and Q.G. analyzed data and co-wrote the manuscript.

## Additional information

**Competing financial interests:** The authors declare no competing financial interests.

# Figure captions

**Figure 1 | Characteristics of the all-optical logic discriminator.** (a) Schematic cross-section structure of a single discriminator unit. (b) Top-view SEM image of the logic discriminator. (c) Calculated SPP propagation length as a function of incident light wavelength for a 300-nm thick gold film coated with a 150-nm thick PVA layer. (d) Calculated magnetic field distribution in a plane 75 nm above the gold film for an 800-nm incident light in the Au/air and Au/PVA/air structures. (e) Calculated power flow distribution in the Y direction 600 nm away from the nanoslit center for an 800-nm incident light in the Au/air and Au/PVA/air structures.

**Figure 2 | Wide-band unidirectional SPP launching properties.** (a) Top-view SEM image. (b) Calculated transmission spectrum of the sinistral 1D plasmonic crystal as a function of incident light wavelength. Measured CCD images of the launcher sample under excitation by a CW laser with wavelengths 750 nm (c), 800 nm (d), 850 nm (e), 900 nm (f), 950 nm (g), and 1000 nm (h). (i) Measured efficiency ratio as a function of incident laser wavelength. (j) Measured CCD image of the launcher sample under excitation by a CW laser at wavelength 532 nm.

**Figure 3 | Performances of the all-optical plasmonic discriminator.** (a) Calculated transmission spectrum of the dextral 1D plasmonic crystal as a function of incident light wavelength for units 1, 2, and 3. PC1, 2, and 3 represent the dextral plasmonic crystals for units 1, 2, and 3, respectively. The dashed lines indicate the position of the stop bandedge in the long-wavelength direction of dextral plasmonic crystals for units 1, 2, and 3. Measured CCD images of the logic discriminator sample under excitation by CW laser at wavelengths 750 nm (b), 800 nm (c), 850 nm (d), 900 nm (e), 950 nm (f), and 1000 nm (g).

**Figure 4** | Calculated power flow distribution in the cross-section of three logic discriminator units under excitation by different CW light signals at wavelengths 750 nm (a), 800 nm (b), 850 nm (c), 900 nm (d), 950 nm (e), and 1000 nm (f).

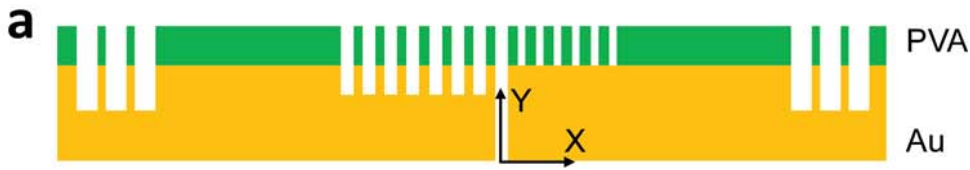

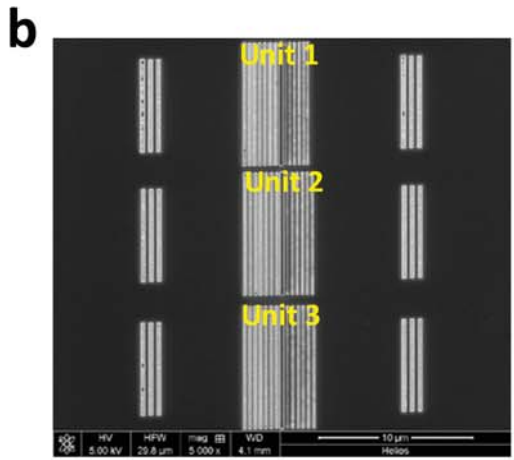

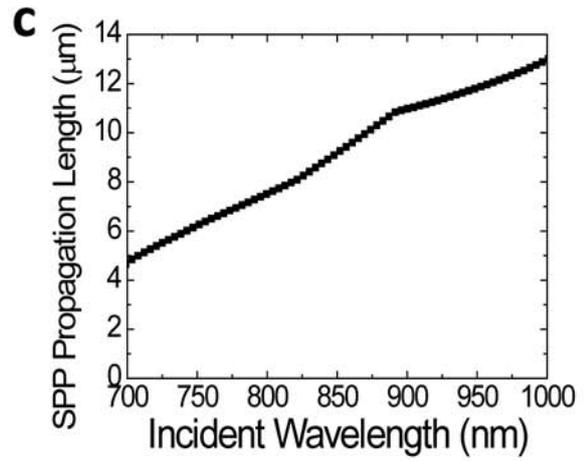

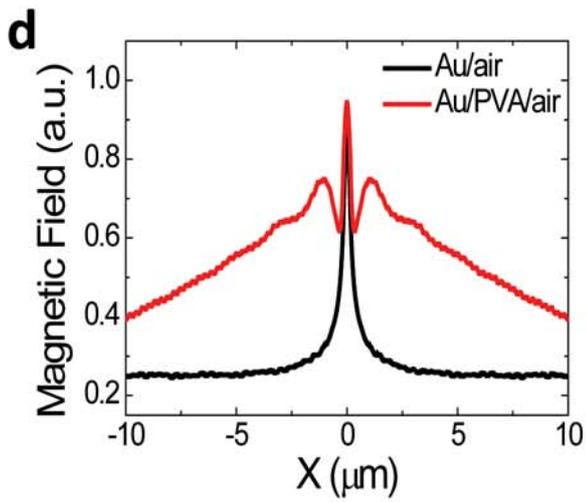

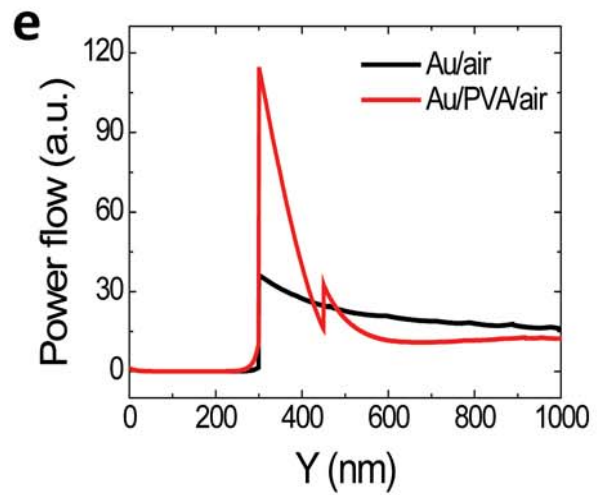

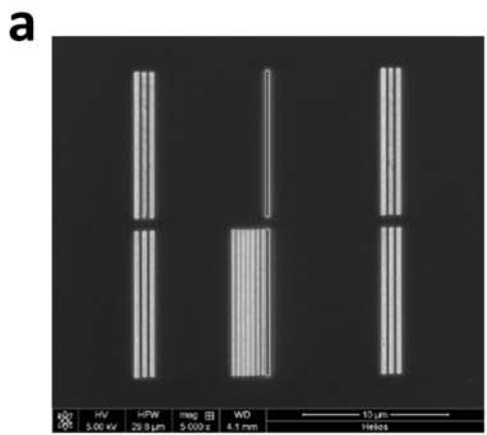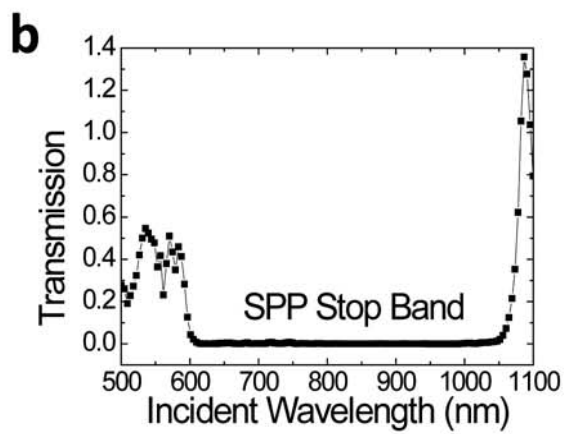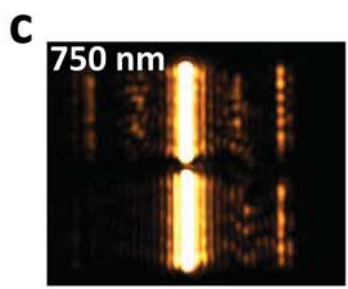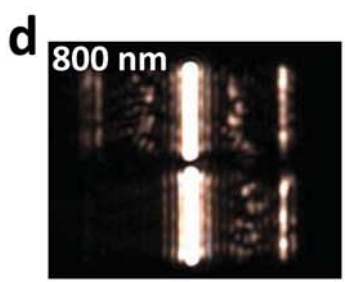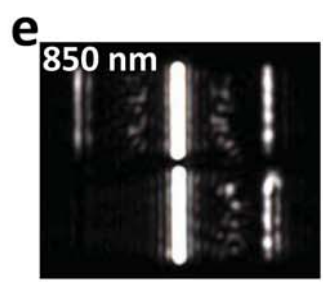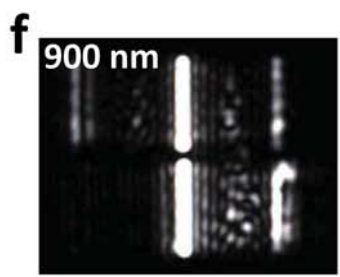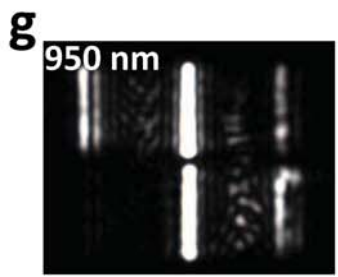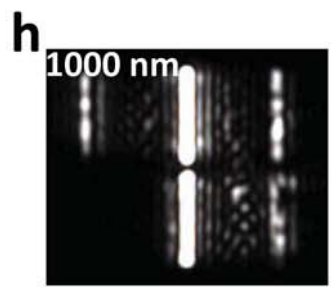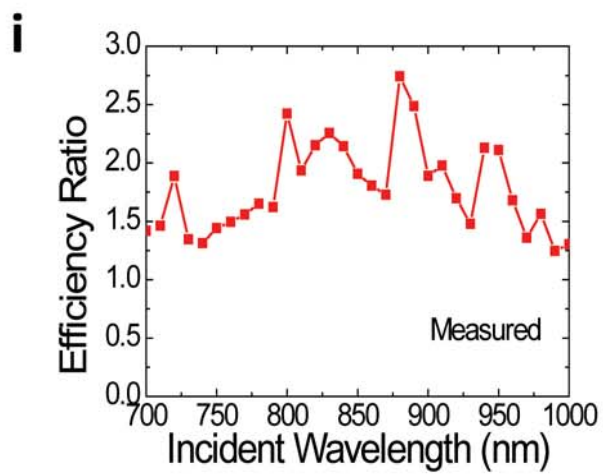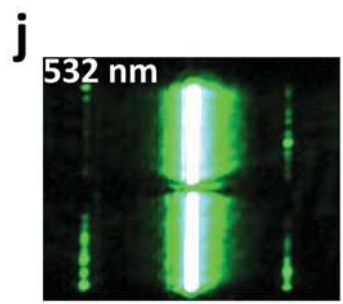

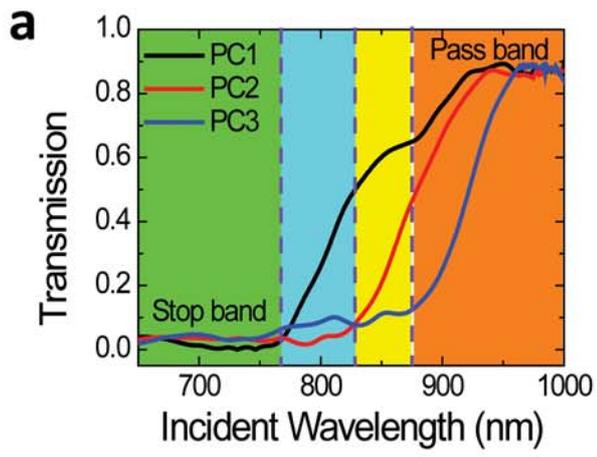
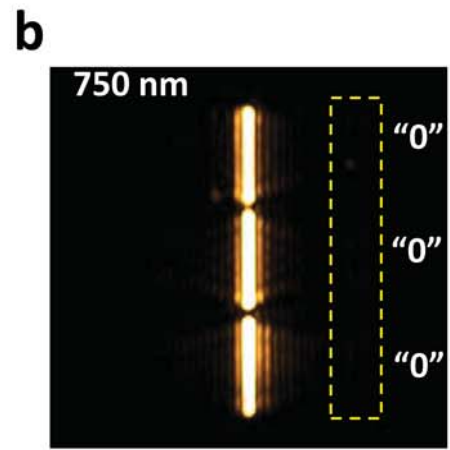
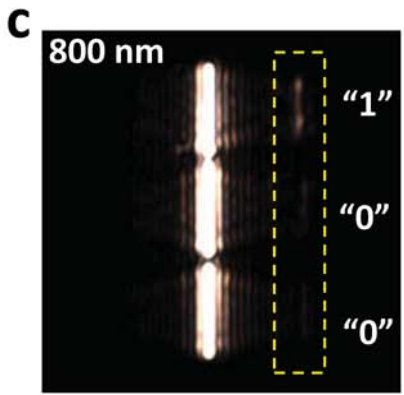
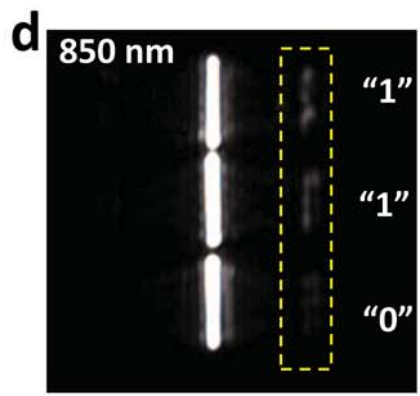
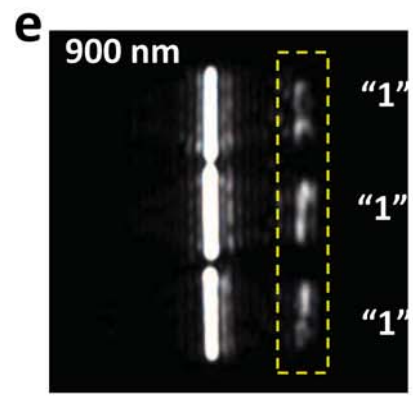
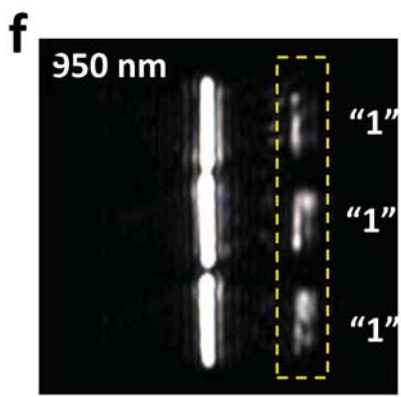
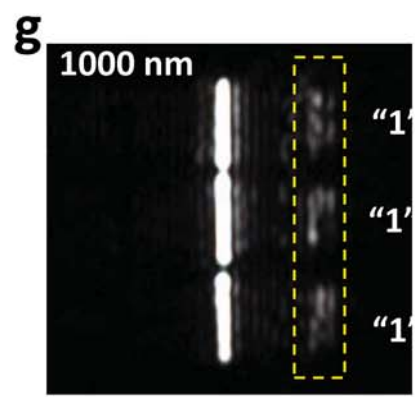

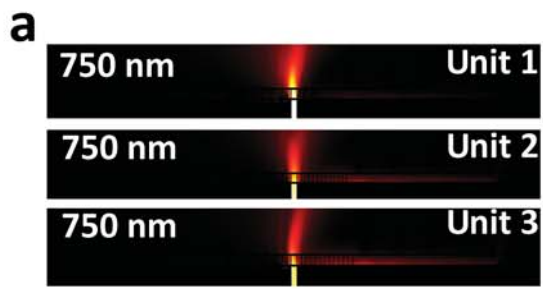
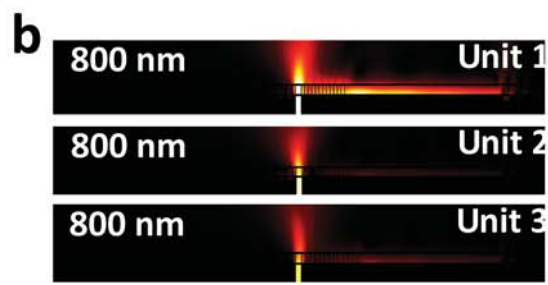
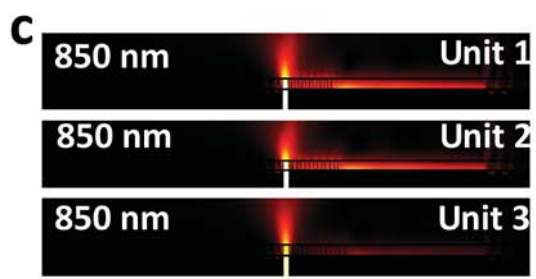
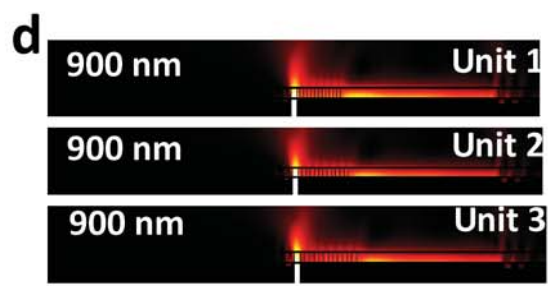
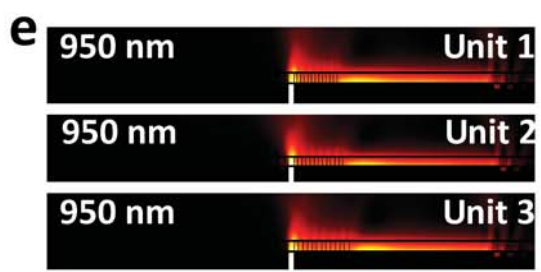
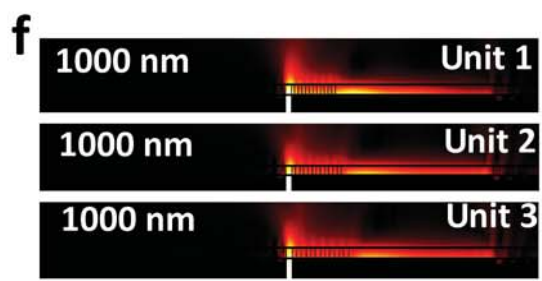